# Historical Seismic Monitoring of EGS and Conventional Geothermal Fields: LBNL Efforts and Lessons Learned

**Nori Nakata[a*], Ernest Majer[a], Chet Hopp[a], Steve Jarpe[b], Tim Elnitiarta[a], Hongrui Qiu[c], Michelle Robertson[a]**

**[a]Lawrence Berkeley National Laboratory, One Cyclotron Road, Berkeley, CA 94720, USA; nnakata@lbl.gov, elmajer@lbl.gov, chopp@lbl.gov, telni1@lbl.gov, MCRobertson@lbl.gov**

**[b]Jarpe Data Solutions Inc., Prescott Valley, AZ 86314; jarpe@pobox.com**

**[c]China University of Geosciences, 388 Lumo Road, Wuhan, P.R. China; qiuhongrui@cug.edu.cn**

*corresponding author: Nori Nakata (nnakata@lbl.gov)

## ABSTRACT

To meet the growing demands and national goals for the deployment of Enhanced Geothermal Systems (EGS) across the United States, it is essential to expand EGS viability across a wide range of geological settings. These include variations in lithology, temperature, subsurface stress regimes, and fluid conditions. One of the most significant challenges in scaling EGS is managing induced seismicity, which can impact both public perception and long-term operational sustainability.

At Lawrence Berkeley National Laboratory (LBNL), we have established and operated seismic monitoring systems across a diverse portfolio of EGS and conventional geothermal sites. These include the Geysers, Desert Peak, Brady Hot Springs, Raft River, and Newberry geothermal fields, where borehole and/or surface seismic arrays have been deployed at each field. In recent years, our monitoring efforts have expanded to additional key sites such as Patua, Don A. Campbell, Jersey Valley, Utah FORGE, and Cape Modern, reflecting a broader national effort to understand seismic behavior across varied reservoir environments.

Through these multi-site deployments, our goal is to better understand the mechanisms underlying induced seismicity and to develop site-specific and transferable best practices for effective seismic monitoring in EGS development. Seismic observations not only provide insights into microseismic activity and reservoir evolution, but also support integrated modeling efforts related to thermo-hydrological-mechanical-chemical (THMC) processes, risk assessment, and reservoir management strategies.

Here, we provide an overview of the status of seismic networks installed by LBNL at the sites listed above and share lessons learned during data acquisition, sensor deployment, and long-term monitoring. We also discuss challenges encountered, such as noise mitigation, sensor coupling, data transmission, real-time processing, and data quality assurance, and the solutions we have adopted.

Finally, we highlight our ongoing efforts to make these valuable seismic datasets accessible to the broader research community. We introduce pathways for data access, aimed at supporting collaboration and advancing further research by scientists, developers, and stakeholders.

**Keywords**

*Seismic monitoring, EGS, Induced seismicity, historical data*

**Highlights**

- Long-term seismic monitoring data recorded by LBNL are reviewed.
- Seismic networks span multiple geothermal fields and EGS projects.
- Publicly accessible datasets support geothermal and induced seismicity research.
- Lessons learned guide seismic monitoring for future EGS developments.

## 1. Introduction

The U.S. Department of Energy (DOE) is investing in Enhanced Geothermal Systems (EGS) as a scalable, baseload renewable energy source to help meet the nation's decarbonization goals. EGS has the potential to unlock geothermal resources in areas without sufficient natural permeability, but its widespread deployment hinges on the ability to monitor and manage induced seismicity—both to mitigate operational risks and to gain public trust (Majer et al., 2016).

This paper summarizes Lawrence Berkeley National Laboratory's (LBNL) extensive seismic monitoring activities across a diverse set of geothermal sites, including EGS demonstrations, conventional, hydrothermal fields, and greenfield explorations. These include The Geysers, Brady's Hot Springs, Desert Peak, Newberry, Raft River, Patua, New York Canyon, Fallon FORGE, Don A. Campbell, Jersey Valley, Utah FORGE, and Cape Modern. At each site, we have installed, operated, and maintained seismic networks tailored to site-specific geological and operational conditions. The networks vary in size and configuration—from borehole geophones and optical accelerometers to surface nodal arrays—and have evolved to improve detection thresholds and spatial resolution.

A major objective of our work is to support reservoir development and management by collecting high-quality microseismic data during stimulation and production. Beyond seismic hazard evaluation, these data provide insights into reservoir structure, stress state, and fracture behavior. At sites like Patua and Desert Peak, we have shown how dense monitoring arrays reduce the magnitude of completeness to below M0.0, enabling the detection of small events linked to localized fracture activation. These observations inform reservoir models and help validate injection strategies in real time.

Another focus of this work is to enhance real-time data processing and transmission. Recent deployments employ robust telemetry—including 3G/4G cellular and satellite communication—to stream high-frequency ($\geq$1 kHz), three-component seismic data to central servers. This facilitates near-instantaneous detection, phase picking, and magnitude estimation through automated processing pipelines. At Jersey Valley and Don A. Campbell, we demonstrate that modern telemetry and adaptive protocols can overcome many limitations of legacy monitoring systems.

This paper also presents lessons learned from over decades of site deployments. These include sensor installation practices, permitting and land access considerations, noise mitigation strategies, and recommendations for calibration and quality control. We discuss how trade-offs between borehole depth, array density, sensor type, and cost affect the performance of seismic networks. Additionally, we emphasize the growing importance of scalable and high-temperature-compatible sensors, especially for long-term monitoring near stimulation zones.

Ultimately, the goal of this work is to contribute to best practices for seismic monitoring of EGS projects, grounded in real-world deployment experience. By integrating engineering needs with scientific insight, and by openly sharing data and methods, we aim to help enable the safe, effective, and publicly acceptable expansion of geothermal energy in the U.S.

## 2. Case studies

LBNL has deployed seismic monitoring systems at geothermal fields across a range of geologic settings, resource types, and development stages in the western United States (Figure 1). These deployments include conventional hydrothermal fields, enhanced geothermal system (EGS) demonstrations, greenfield exploration projects, and research test sites. Together, they provide a broad basis for evaluating how seismic network design, instrumentation, telemetry, and data-processing strategies vary with project goals and site conditions.

In this section, we summarize the monitoring history and network characteristics for the principal sites discussed in this paper. Sensor specifications and monitoring durations are listed in Table 1, and the following subsections describe the deployment context, network evolution, and major monitoring objectives at each geothermal site.

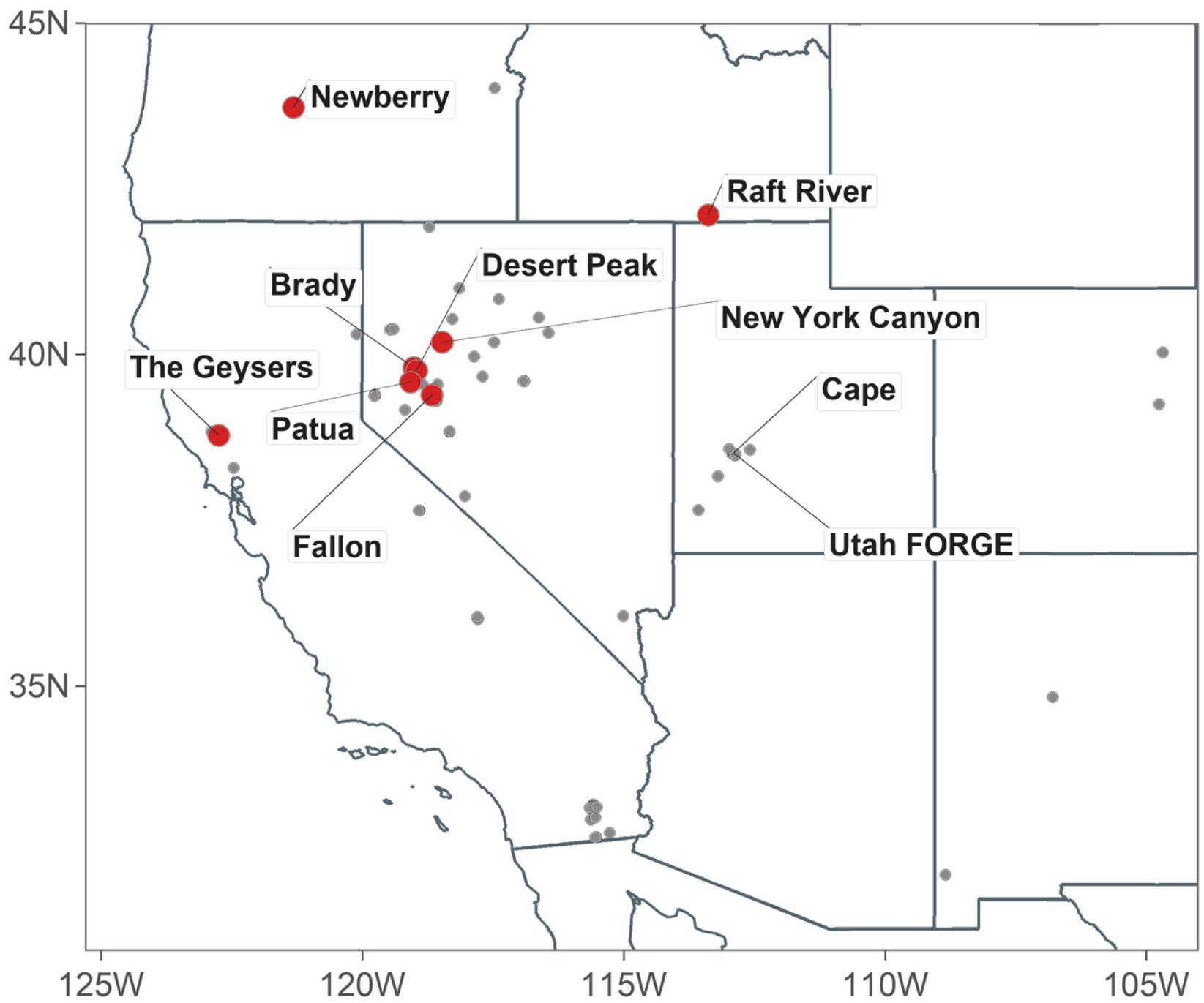

**Figure 1.** Map of geothermal sites in the western United States. Red circles indicate sites where LBNL has deployed seismic monitoring systems and that are discussed in this study; gray circles show other operating or developing geothermal fields. Utah FORGE and Cape are labeled for geographic reference. Geothermal field locations are from the Global Energy Monitor Global Geothermal Power Tracker (March 2026).

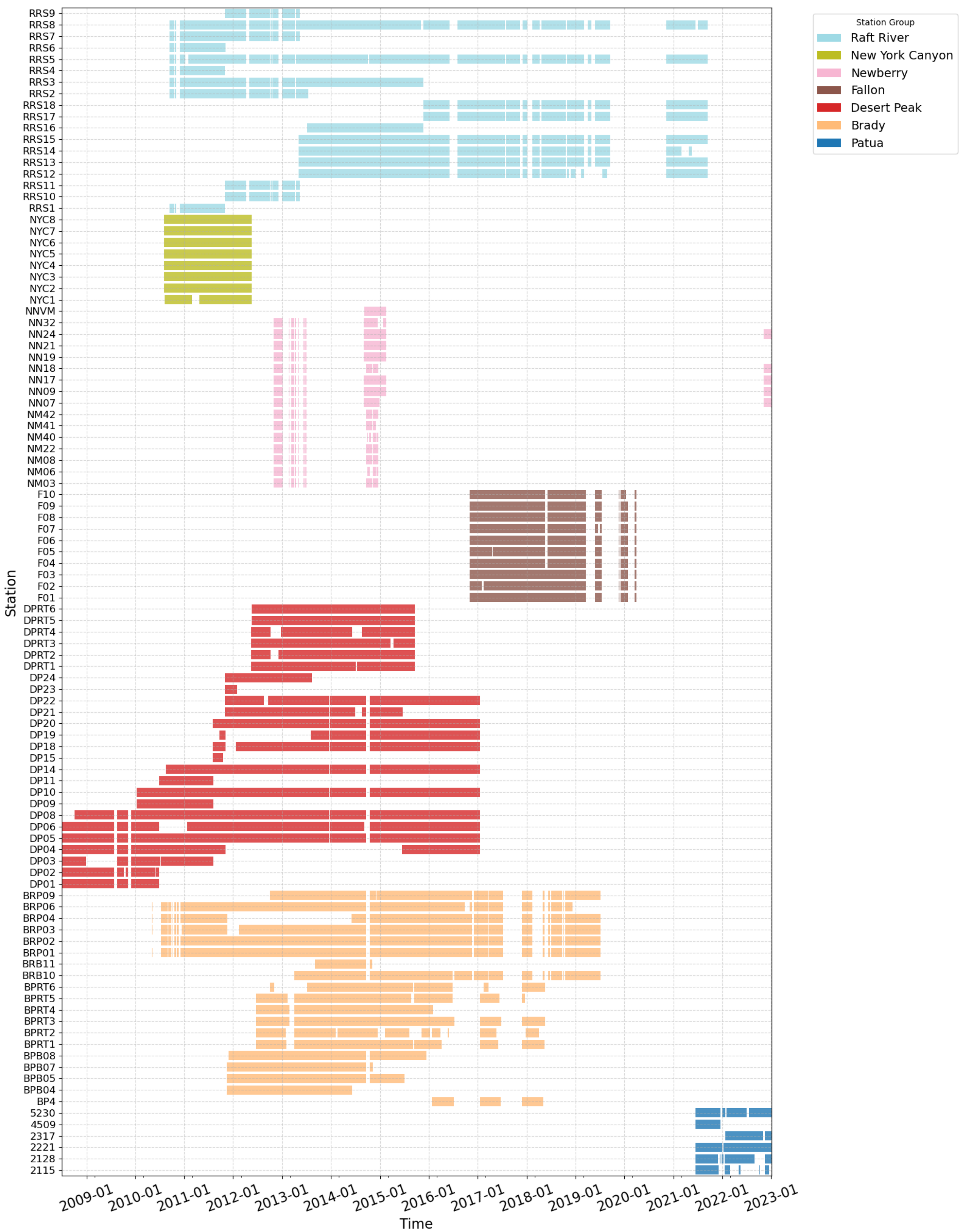

**Figure 2:** Data availability plot for the networks described in this work. Bars are condensed into weekly bins for ease of visualization. Geysers data are omitted.

### *2.1 Catalog generation for case studies*

The seismic monitoring systems discussed in this paper were deployed at different times, with different telemetry configurations, station geometries, and project objectives. As a result, the case-study catalogs were generated using site-specific workflows rather than a single uniform processing system. The Geysers (early years), Desert Peak, Brady's Hot Springs, Raft River, Newberry (early years), and Patua (early years) datasets were recorded as event gathers identified by computers running at each site. The event detection process computed an STA/LTA value for each station from the continuous data, then used a minimum number of triggered stations within a time window to determine if a possible event occurred. For each declared event, a 30-s time window of data from all stations was gathered and transmitted to LBNL. For other sites, continuous waveform data were transmitted to LBNL, where event detection was performed.

For each event gather, preliminary P-wave arrival times were determined using an STA/LTA-based approach. P-wave arrivals from all stations were examined, and events were retained when approximately four to five arrivals occurred within a narrow time window of about 3 s. A 30-s event gather could contain zero, one, or multiple events. For each retained event, the P-wave arrival was used as a starting point to identify S-wave arrivals using an STA/LTA-based approach. Event origin times and hypocenter locations were determined from the P- and S-wave arrivals using SimulPS. (Evans et al., 1994) A three-dimensional velocity model was used for The Geysers, whereas one-dimensional velocity models were used for the other sites. Moment magnitudes were estimated from spectra of waveform windows that included the P- and S-wave arrivals at each station, and the event magnitude was calculated as the average of the station magnitudes.

The final Geysers catalog resulted from the event-based processing workflow. For Desert Peak, Brady's Hot Springs, Raft River, Patua, and Newberry, all events were manually reviewed; regional earthquake and non-earthquake events were removed, arrivals were repicked where needed, and events were relocated to produce the final catalogs used in the case studies.

**Table 1.** Seismic monitoring specification for each site with injection activities, if present.

| Site | Seismic monitoring term | Seismic stations | EGS Injection | FDSN Network | Continuous or triggered | Key publications |
|---|---|---|---|---|---|---|
| The Geysers | 1974-present | (34) 3C, 4.5 Hz 500 s/sec<br>(5) 3C, 2.0 Hz, 500 s/sec<br>(100) 3C, 4.5 Hz<br>(1) strong motion | 2011 | BG | Triggered (1974-2013), Continuous (2013-) | Majer and Peterson (2007)<br>Garcia et al. (2016)<br>Rutqvist et al. (2016) |
| Desert Peak | 2008-2016 | (10) 3C, 4.5Hz, 500 s/sec (3 borehole)<br>(6) 3C, 2.0 Hz | 2010-11, 2013 | 5I | Triggered | |
| Brady's Hot Springs | 2010-2019 | (8) 3C, 8.0 Hz, 500 s/sec (5 borehole)<br>(6) 3C, 2.0 Hz | 2011, 2013 | 4I | Triggered | Akerley et al. (2021) |
| Raft River | 2010-2022 | (9) 3C, 4.5 Hz, 500 s/sec (3 borehole)<br>(15) 3C, 4.5 Hz | 2011 | 6I | Triggered | Bradford et al. (2017) |
| Newberry | 2010-present | (8) 3C, 4.5 Hz, 500 s/sec (7 borehole)<br>(1) strong motion<br>(~20) 3C, 5.0 Hz, 1000 s/sec | 2012, 2014, 2024 | 9G | Triggered (2010-2021), Continuous (2021-) | Cladouhos et al. (2016)<br>Templeton et al. (2020)<br>Sonnenthal et al. (2015)<br>Nakata et al. (2026) |
| New York Canyon | 2011-2012 | (8) 3C, 4.5 Hz, 500 s/sec, borehole | | 6B | Continuous | |
| Patua | 2012-2013<br>2021-2023 | (6) 3C, 15 Hz, borehole (18 stations in 2012-2013)<br>(6) 3C, 2.0 Hz, surface | | 24 | Triggered (2012-2013), Continuous (2021-2023) | Nakata et al. (2023) |
| Fallon FORGE | 2016-2020 | (6) 3C, 4.5 Hz, 250 s/sec, borehole<br>(4) 3C, 8 Hz, 250 s/sec, borehole | | 3W | Continuous | |
| Cape Modern & Utah FORGE | 2023-present | (6) 3C, 4.5 Hz, borehole<br>(500-1000) 3C, 5.0 Hz, surface | 2019, 2022, 2024- | 6K | Continuous | Nakata et al. (2024)<br>Nakata et al. (2025) |

### *2.2 The Geysers*

The Berkeley Geysers (BG) network has operated since 1989, and seismic monitoring at The Geysers dates back to the 1960s (Majer 1978). Since 2003, 40 BG stations have been installed, and the current monitoring system also includes seven stations from the Northern California (NC) network. This seismic network provides an approximate magnitude of completeness of 0.8 across the Geysers reservoir (Figure 3). The Geysers network is currently managed by Calpine, and waveform data are available through the Northern California Earthquake Data Center (NCEDC) under network code BG. Seismic data are analyzed by the USGS, and the resulting earthquake catalog is integrated into the USGS catalog.

Within The Geysers, the Northwest Geysers EGS Demonstration Project provides a key example of how integrated seismic monitoring can be used to evaluate reservoir stimulation in a complex geothermal setting (Garcia et al., 2016; Rutqvist et al., 2016). In this project, the abandoned Prati 32 and Prati State 31 wells were reopened and deepened to access a high-temperature reservoir beneath the conventional Geysers steam reservoir, and treated wastewater injection into P-32 began in October 2011. Microseismic monitoring using the permanent Geysers network and temporary focused arrays tracked the growth of the stimulated volume, identified preferential flow paths along permeable shear zones, and evaluated the response of induced seismicity to injection-rate changes. Together with coupled thermal-hydraulic-mechanical modeling, these observations showed that stimulation was controlled by both pressure changes and cooling-induced stress changes, demonstrating how seismic datasets can constrain reservoir structure, stimulation extent, and operational response in EGS projects.

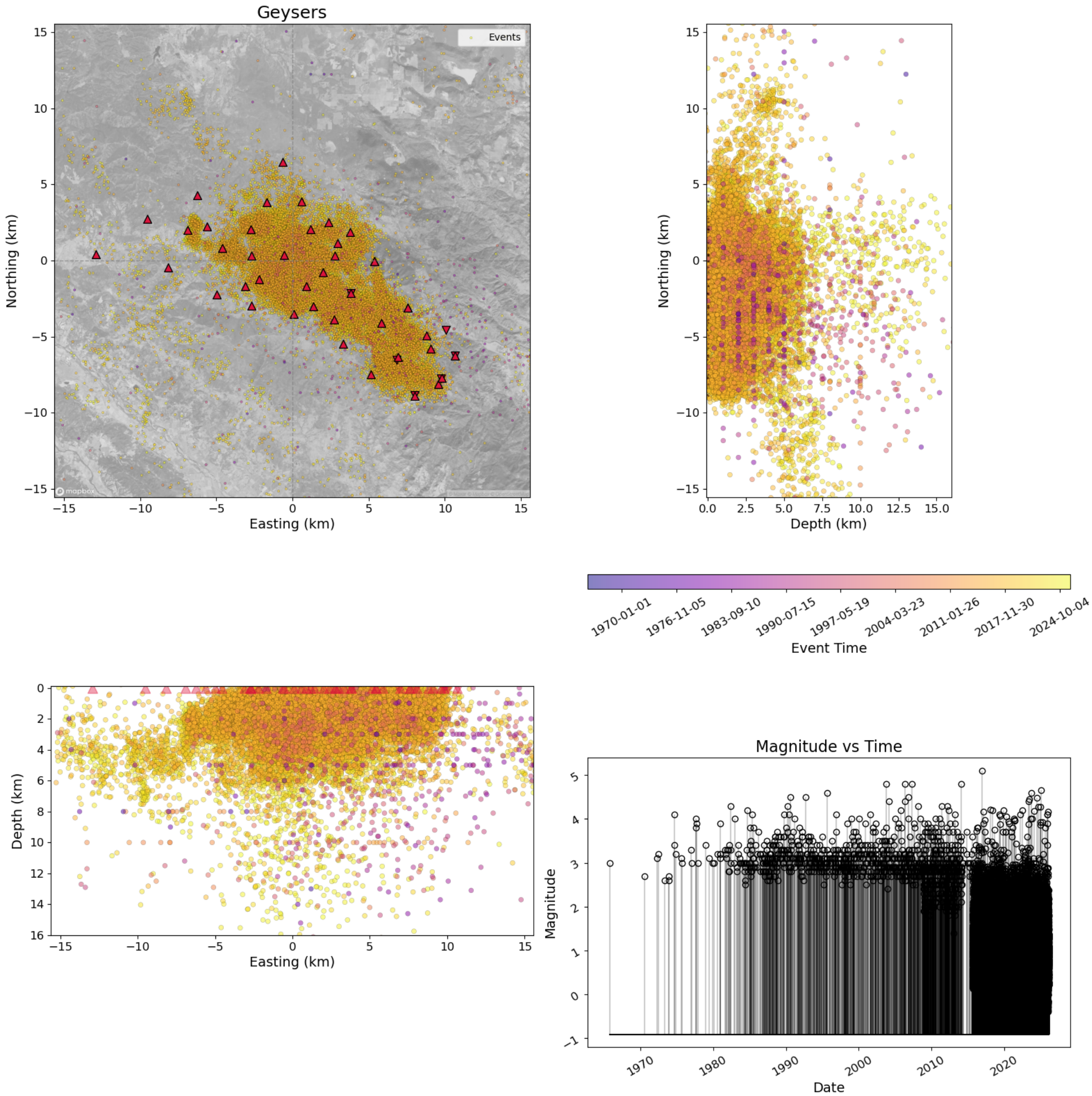

**Figure 3.** 124,532 events have been recorded at the Geysers geothermal field since 1965 (as retrieved from the SAGE FDSN services on 1-23-2026). Stations from both the NC and BG networks are shown as red triangles (inverted for stations installed in boreholes). Events are colored by date of occurrence. The magnitude of the events with time is shown in the lower right.

## *2.3 Desert Peak*

In 2010, a 14-sensor network was jointly installed by LBNL and USGS to monitor the EGS activities at the Desert Peak geothermal field in Churchill County, NV. In 2011, the network was expanded with an additional four boreholes at a depth of 300 ft. The network was in operation from 2008 until 2017, spanning the injection activities that lasted from September 2010 until

March 2013 (Figure 4; Ackerly et al., 2021; Chabora et al., 2012). Desert Peak is a producing hydrothermal field where the EGS effort targeted permeability enhancement in and around an existing reservoir volume rather than a greenfield resource. The seismic array was designed to detect small induced events during injection and to resolve the spatial relationship between microseismicity, borehole locations, and the stimulated zone.

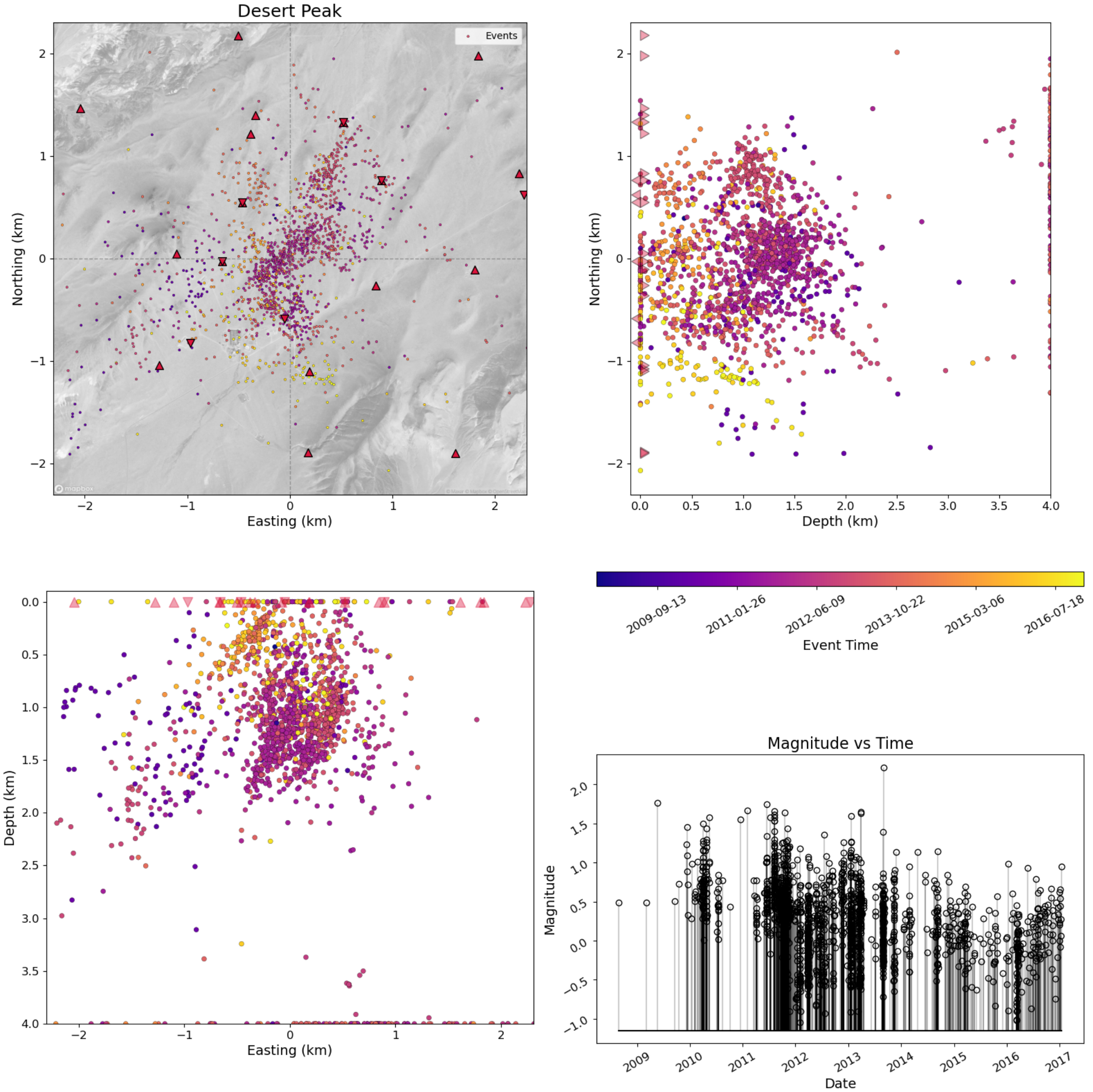

**Figure 4.** Seismicity and the seismic array at Desert Peak. Seismic stations are shown as red triangles (inverted for stations installed in boreholes). Events are colored by date of occurrence. The magnitude of the events with time is shown in the lower right.

### *2.4 Brady's Hot Springs*

Following the injections at Desert Peak, LBNL installed six 300 ft-deep 3C borehole sensors and three surface sensors to monitor EGS injection activities at Brady Geothermal Field, approximately 5 miles northwest of Desert Peak in Churchill County, NV. The network recorded from 2010-2019 (some sensors), with the main focus being the April-October 2013 injection activities at the field (Figure 5; Akerley et al., 2021). Brady provides a close comparison to Desert Peak because both fields are located in the same general region but were monitored during separate stimulation campaigns. The mixed borehole and surface array was intended to improve detection of small events and constrain the spatial relationship between induced seismicity, shallow structures, and injection operations.

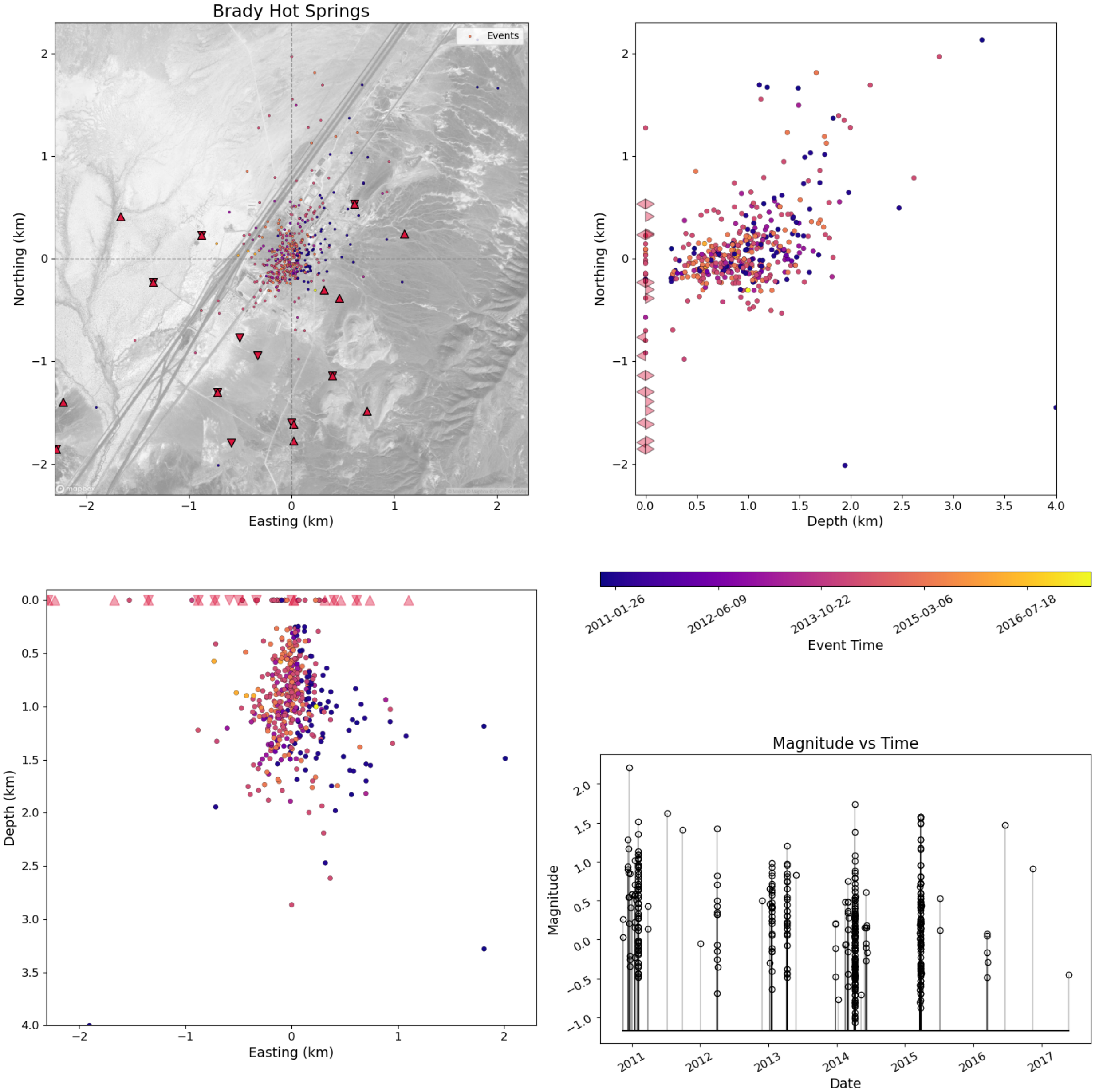

**Figure 5.** Seismicity and the seismic array at Brady Hot Springs. Seismic stations are shown as red triangles (inverted for stations installed in boreholes). Events are colored by date of occurrence. The magnitude of the events with time is shown in the lower right.

### *2.5 Raft River*

The Raft River seismic network was installed beginning in August 2010 and removed in November 2022. For most of its life, it comprised four 100-ft deep borehole geophones and four surface geophones and was installed to monitor the EGS demonstration project that included multiple stimulations from February 2012 through the end of 2013 (Figure 6; Bradford et al., 2017; Li et al., 2017). Raft River provides a longer-term example of monitoring around an EGS demonstration embedded within an operating geothermal field. The combination of shallow borehole and surface sensors allowed the catalog to capture both near-well seismicity associated with stimulation and broader wellfield-scale seismic response over multiple years.

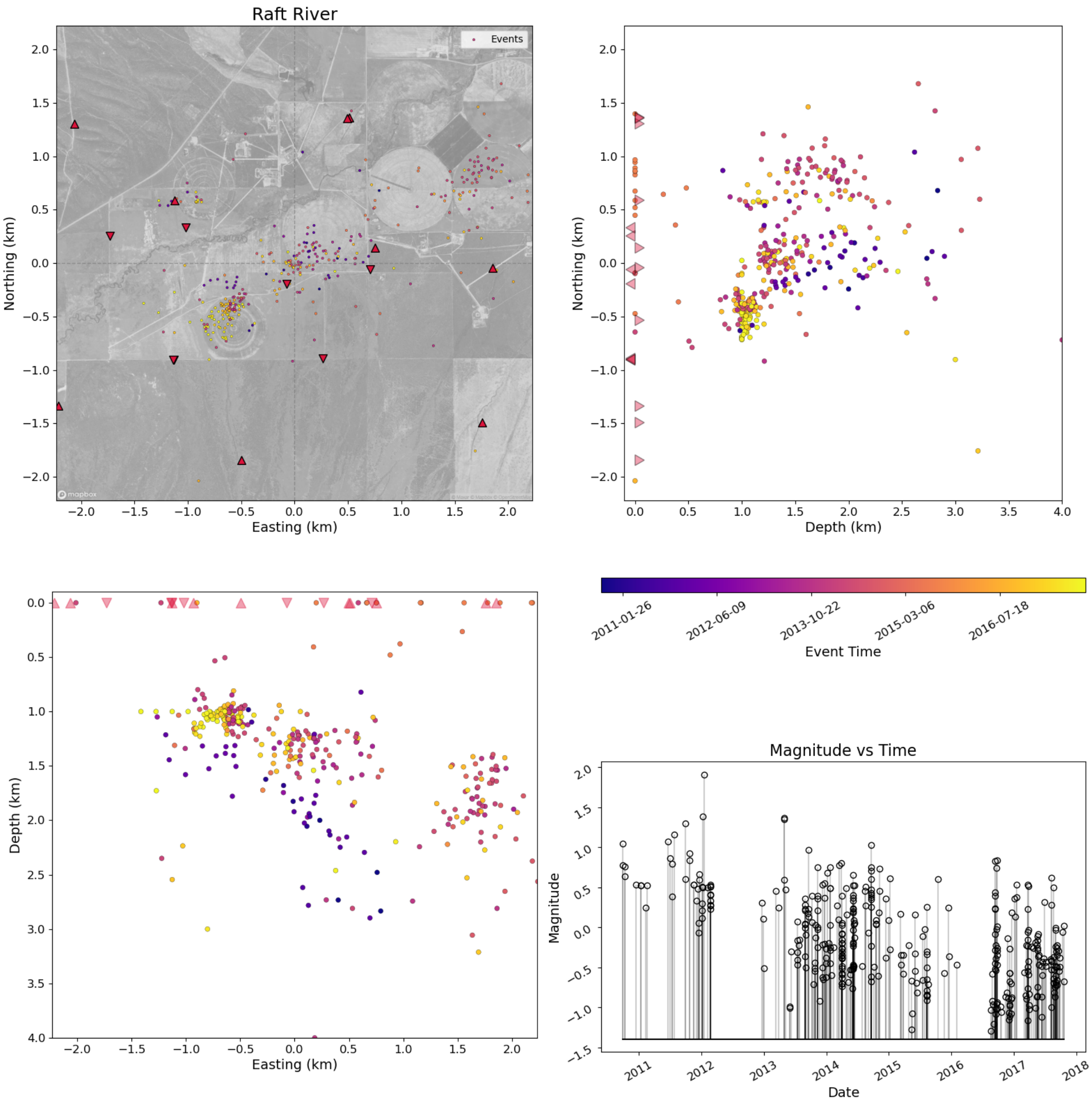

**Figure 6.** Seismicity and the seismic array at Raft River. Seismic stations are shown as red triangles (inverted for stations installed in boreholes). Events are colored by date of occurrence. The magnitude of the events with time is shown in the lower right.

### *2.6 Patua Geothermal site*

Between July 2012 and late 2013, an array of 18 borehole 15-Hz geophones operated at Patua. While most equipment was removed afterward, many sensors remained grouted or locked in place. In June 2021, LBNL began reinstalling recorders and testing these geophones. A 2-Hz surface geophone was also added at each site to assess sediment attenuation. From 2021-2023, data were recorded from six of the original borehole sensors (Figure 7).

The Patua network aimed to detect and locate as many earthquakes as possible to reduce seismic risk and map active structures that are essential to reservoir flow modeling. With all 18 borehole sensors, the network achieved a completeness magnitude of ~0.0, capturing events linked to fractures as small as 5–10 meters (Nakata et al., 2023). Figure 7 shows only events detected from 2021–2023 by up to six stations. Earlier data is currently unavailable due to a non-disclosure agreement.

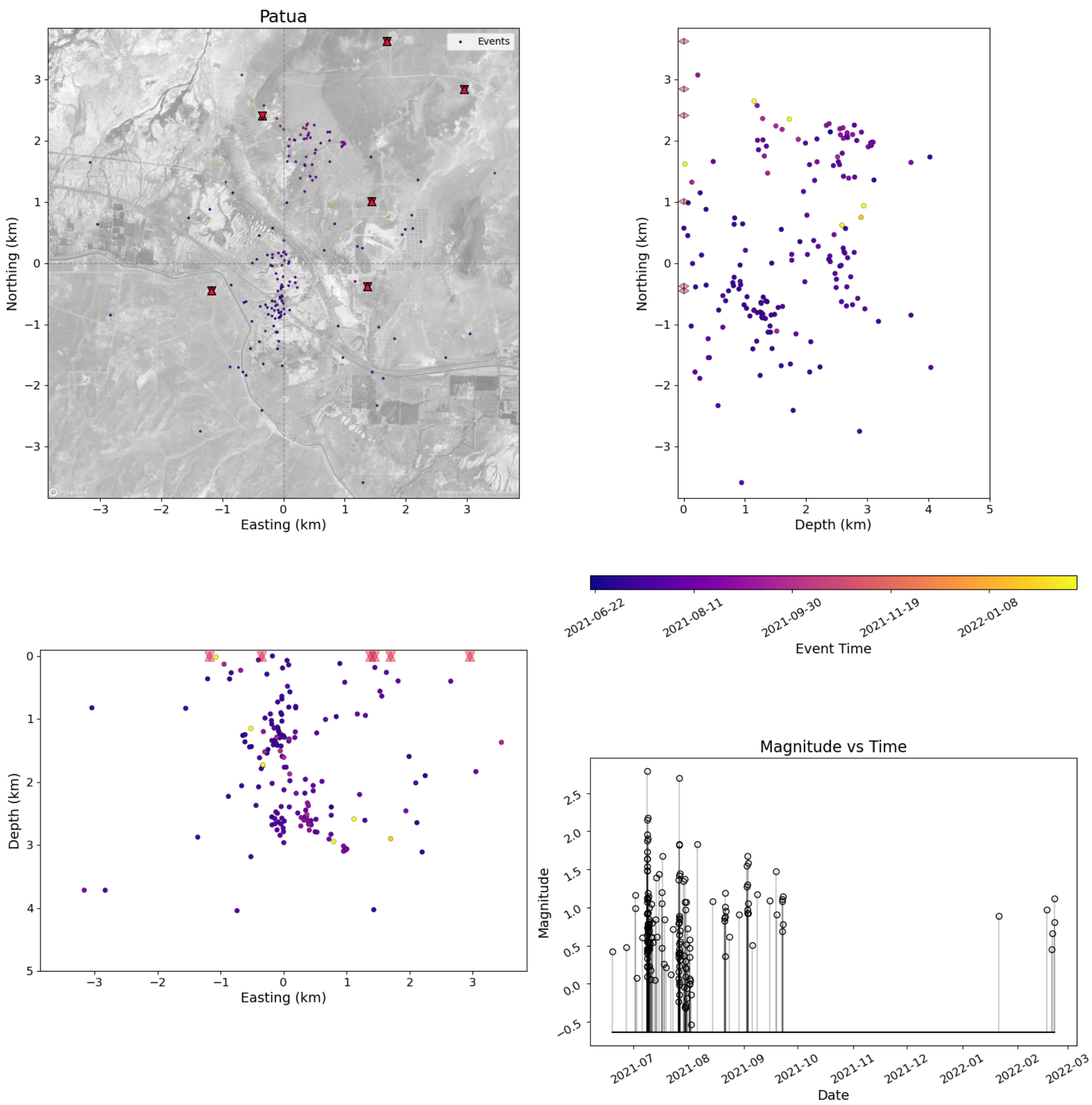

**Figure 7.** Seismicity and the seismic array at Patua from 2021–2023. The seismicity during the earlier deployment (2012–2013) is not shown due to an NDA. Seismic stations are shown as red triangles (inverted for stations installed in boreholes). Events are colored by date of occurrence. The magnitude of the events with time is shown in the lower right.

### *2.7 Newberry*

The seismic monitoring system currently deployed at the Mazama Newberry EGS site closely resembles the configuration used during AltaRock's earlier projects in 2012 and 2014 (FDSN code 9G; Sonnenthal et al., 2015; Cladouhos et al., 2016). It includes eight sensors placed in boreholes reaching depths of up to 300 meters. Of these, six (NN07, NN09, NN17, NN19, NN21, and NN24) are the original 2 Hz geophones since 2012, while the remaining two (NN18 and NN32) are broadband accelerometers added in 2023. The collected data is streamed nearly in real time to NSF SAGE, where it is publicly accessible. Newberry is a volcanic EGS site, and the current monitoring system builds on earlier stimulation experiments while supporting renewed injection and real-time seismicity analysis. The combination of local borehole sensors and regional stations is important for distinguishing injection-related microseismicity from broader background or volcanic seismicity near the Newberry system.

Additional seismic coverage is provided by nearby stations operated by the University of Washington's Pacific Northwest Seismic Network (PNSN) and the USGS Cascades Volcano Observatory (CVO).

All data from these stations is processed in real time using LBNL's detection and analysis system, which identifies, locates, and estimates magnitudes of seismic events related to the ongoing injection activities at the Newberry site (Figure 8).

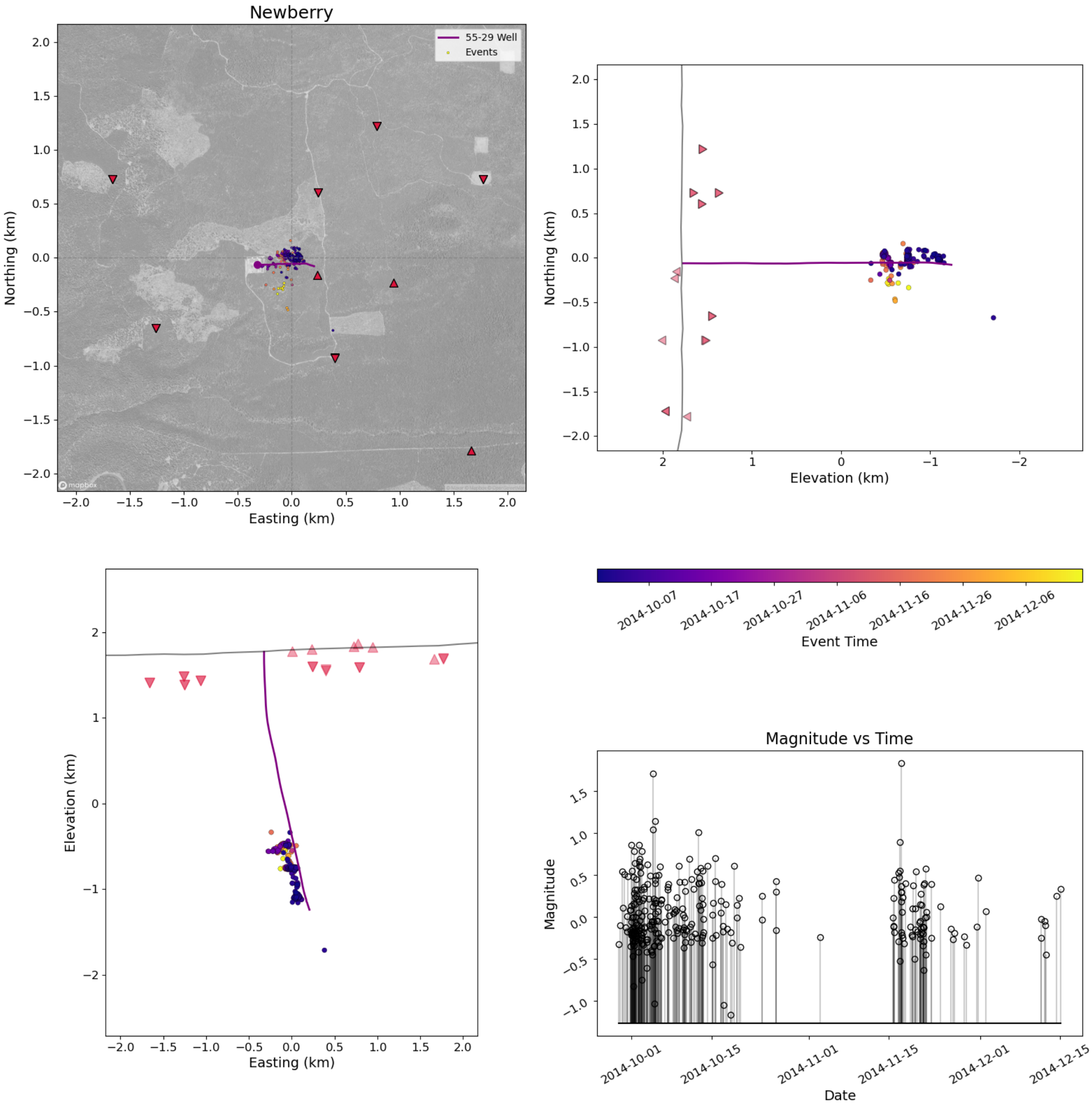

**Figure 8.** Seismicity and the seismic array at Newberry. Seismic stations are shown as red triangles (inverted for stations installed in boreholes). Events are colored by date of occurrence. The magnitude of the events with time is shown in the lower right. The well stimulated during the EGS operations from 2012 through 2014 (55-29) is shown in purple.

### *2.8 Greenfield exploration sites*

LBNL also deployed two networks for initial site characterization at New York Canyon and Fallon FORGE. These deployments represent greenfield exploration settings, where short-term seismic monitoring can establish baseline seismicity and test whether a proposed resource area has detectable local activity before larger-scale development. The New York Canyon array (FDSN code 6B) included eight three-component geophones deployed at the surface from late 2011 until mid 2012. Processing of the data revealed no local seismic events and the array was decommissioned. The Fallon FORGE array (FDSN code 3W) was in operation from 2016 to 2020 in support of the initial phases of the DOE FORGE program. As with New York Canyon, few local events were detected during the Fallon FORGE deployment, providing useful baseline information for interpreting later exploration or stimulation activities.

### *2.9 Modern seismic array*

Modern seismic arrays often combine several complementary sensor types, including dense surface nodes with or without telemetry, deep borehole high-temperature seismometers, and deep-borehole distributed acoustic sensing (DAS), in addition to the backbone surface or shallow-borehole seismometers discussed in the previous sections. These deployments are designed to improve spatial resolution, capture small events close to stimulation zones, understand structure changes, and test instrumentation that can operate in high-temperature geothermal environments (e.g., Chang et al., 2022). For example, at the Cape Modern and Utah FORGE geothermal field area, we installed six permanent seismic monitoring stations and approximately 500–1000 temporary nodal sensors to monitor seismic activity, complementing the extensive preexisting network operated by the University of Utah Seismograph Stations. The temporary nodal array primarily consists of SmartSolo IGU-16HR 3C 5 Hz sensors. Some of these nodal sensors operate without telemetry and are therefore used primarily for retrospective detailed analysis, while telemetered stations support near-real-time monitoring. Additionally, twelve SmartSolo IMU stations equipped with telemetry were deployed, totaling 408 geophones grouped to increase the effective network density. In addition to these surface and near-surface sensors, one to three deep-borehole DAS systems and one high-sensitivity, high-temperature borehole seismometer installed at a depth of 2132 m provide complementary subsurface monitoring (Nakata et al., 2024; 2025; 2026b).

### *2.10 Data accessibility*

Waveform data for the networks detailed above can be accessed through FDSN web services from NSF SAGE. These networks have been consolidated under the SAGE virtual network code "_LBNL," alongside contemporary LBNL networks currently in service. The individual networks listed in this paper can be accessed by their FDSN temporary network codes summarized in Table 1. For most sites, data were recorded in triggered mode, meaning that the archive is not continuous but instead consists of waveform windows surrounding detected seismic events. In some cases, waveforms are also available through NCEDC, although the NCEDC holdings are not as complete as the SAGE archive. The Geysers network (BG) waveform data are available via FDSN from the NCEDC. The Geysers earthquake catalog is also available from the NCEDC FDSN webservices.

## 3. Automatic processing system

Regional seismic networks routinely process waveform data and publish earthquake catalogs through agencies such as the USGS. These catalogs provide an essential public record, but they are generally optimized for regional earthquake monitoring rather than site-specific geothermal applications, where lower magnitude completeness, consistent processing across deployments, and detailed analysis near injection or production zones are often required. The case-study catalogs described above were generated using site-specific workflows that evolved over multiple deployments. More recently, LBNL has developed an automatic seismicity-processing system to support more consistent processing of geothermal seismic datasets. Incorporating machine learning is one of the recent tasks to speed up the processing and increase the accuracy (Nakata et al., 2025; Bi et al., 2025)

Recently, LBNL has set up an automatic processing system of seismicity using the background seismic network using *SeisComP* (Helmholtz-Centre Postdom, 2008). We then refine the automatically processed earthquake catalog using other datasets. The seismic data processing pipeline for the LBNL SeisComP server comprises the following parts:

1. Data telemetry over the cellular network using the RTP and seedlink protocols

2. Initial detection is conducted using a short-term average/long-term average (STA/LTA) detection algorithm. The detection statistic is calculated on vertical component data bandpass filtered between 3 and 30 Hz with a short-term window of 0.1 seconds and a long-term window of 5 seconds. When the short-term/long-term ratio exceeds 3, a detection is triggered. The detection is turned off once the ratio falls below 1.5. Parameters would be adjusted at each site.

3. Once a detection is triggered, SeisComP launches both a P and S-phase post-picking algorithm to refine the arrival times. Both of these algorithms are based on the Akaike Information Criterion (AIC; Akaike, 1973).

4. All phase picks made by steps 2 and 3 are fed to a phase association algorithm. The phase association is a cluster search based on the DBSCAN algorithm (Ester et al., 1996), whereby P-picks are continually clustered and associated. The clustering metric is the vector sum of the time difference between any pair of picks and the travel time between the stations corresponding to these picks (assuming some average crustal velocity). Once a cluster of P-picks has been formed, additional picks can be associated with this origin from later picks and other origins.

5. Each final origin created in step 4 is then relocated using a linearized location algorithm with a 1D velocity model tailored to the region in which the picks were made.

6. If an initial location falls within a set of predefined regions (within 25 km of an injection well, for example), the origin is then relocated using the NonLinLoc software package with a 3D velocity model (Lomax et al., 2009).

7. Two types of magnitudes are typically estimated: local magnitude (Ml) and duration magnitude (Md). For each pick made, an Ml and Md station magnitude is estimated (Ml being amplitude-based and Md being based on the duration of the event waveform). The network magnitude for each type is computed as the average of all station magnitudes, ignoring outliers beyond the upper and lower 12.5 percentile.

## 4. Lessons Learned from Seismic Monitoring of EGS and Conventional Geothermal Fields

LBNL has been actively monitoring and studying microearthquake (MEQ) activity in geothermal fields since the 1970s, starting with The Geysers and expanding to other regions, including Nevada (Majer 1978; Majer and McEvilly, 1980). Over time, LBNL contributed to international efforts, leading to the development of the "Protocol for Addressing Induced Seismicity Associated with Enhanced Geothermal Systems" (Majer et al., 2009, 2012) and the subsequent "Best Practices" documents (Majer et al., 2012, 2016, 2025). These guidelines, although not state-of-the-science, provide practical strategies grounded in proven methods for hazard mitigation and reservoir management. They are updated as new technologies and insights emerge and are especially valuable for public engagement and regulatory compliance (see Steps 2, 4, and 7 of the Best Practices).

LBNL's work generally serves two purposes: (1) collecting and analyzing MEQ data to assess and mitigate induced seismicity risks while promoting public acceptance, and (2) leveraging MEQ data to optimize reservoir performance using advanced seismic instrumentation.

### *4.1 Hazard and Risk Mitigation*

Experience has shown that proactive outreach is critical—engaging stakeholders before, during, and after drilling can greatly influence project success. Transparent communication, including a balanced explanation of risks and benefits, can convert skeptical stakeholders into supporters. Clear descriptions of mitigation strategies—ranging from engineered approaches (e.g., injection rate adjustments using Traffic Light Systems) to adaptive frameworks (e.g., Adaptive Traffic Light Systems or ATLS; Zhou et al., 2024)—are key to building trust. Indirect mitigation, such as managing liability or working with regulators, is equally important. The acceptable risk level will vary by project and should be defined with operators, regulators, and local communities before operations begin.

### *4.2 Seismic Data Collection for Reservoir Management*

Following the stimulation at Desert Peak, it was determined that the magnitude of completeness of the catalog was too high to capture the smaller, more numerous events needed to illuminate reservoir structure in the required detail (Ackerly et al., 2021). Since 2011, LBNL has increasingly deployed seismic sensors in boreholes to reduce noise and detect smaller magnitude events, a necessity for illuminating reservoir structures at high resolution. Achieving a magnitude of completeness below M0.0 is often critical for reservoir management. Lowering the magnitude of completeness by one magnitude unit can yield roughly an order of magnitude more events, which improves the statistical basis for characterizing reservoir structure and behavior.

In parallel, dense arrays of surface nodal seismometers and geophone strings wired in series have improved detection sensitivity and spatial resolution while reducing the deployment cost with drilling. These configurations often rival borehole quality, especially when subsets of the nodes are streamed in real time to enhance monitoring density. Note that the detectability is different from locatability or characterizability of earthquakes. For locating and characterizing earthquakes, we need denser networks.

Today, data streaming is no longer constrained by telemetry bandwidth for most of the sites in the US; even 3G networks can stream 1 kHz three-component data. When cellular connectivity is unreliable, satellite telemetry offers a robust, cost-effective alternative.

An effective MEQ monitoring system should provide:

- Accurate event locations (x, y, z ± 0.5 km, t ± 0.1 s). Relative locating methods such as HypoDD can increase the accuracy relatively.
- Broad magnitude coverage (from M–2.0 to M+4.5 or more)
- Focal mechanisms (even without full moment tensors)
- Seismicity rates (e.g., Gutenberg-Richter statistics)
- Real-time data access for all stakeholders

For engineering applications, real-time, high-resolution seismic data can validate fluid injection strategies and guide operations. Monitoring systems should ideally digitize at ≥500 Hz with 24-bit resolution and 120 dB dynamic range. A good P- and S-wave velocity model is essential for accurate location and interpretation. If we are interested in the stress changes due to earthquakes, which requires the earthquake corner frequency estimation (Aki, 1967), we need higher sampling rates.

### *4.3 Network Design Considerations*

Designing seismic arrays involves balancing cost, sensitivity, and coverage. Borehole sensors offer high sensitivity but limited spatial coverage; surface stations offer broader coverage but higher noise. Advances in fiber-optic sensing and downhole geophones enable deployment in higher-temperature zones (>225°C), though fully capable three-component broadband sensors (>0.01–1000 Hz, 150 dB dynamic range) remain a pressing need for long-term deployment in 3” boreholes. Recent development of high-temperature borehole seismometers (Nakata et al., 2026b) can lower the magnitude of completeness by about two magnitude units relative to surface sensors, enabling the detection of subtle failure mechanisms (Modes I, II, and III) near stimulation zones.

Land access and permitting also influence station placement—surface stations are usually easy to permit, but borehole installations on public lands (e.g., BLM, USFS) may take months. Real-time telemetry, power supply, and terrain accessibility are other key constraints that must be considered in early planning stages.

### *4.4 Sensor Calibration and Verification*

Prior to operation, calibrating the seismic network is essential. This includes checking sensor polarity and confirming timing consistency across stations using known sources (e.g., explosions), side-by-side comparisons, or large regional events with well-constrained ground motion. Such procedures are critical for accurate focal mechanism and moment tensor inversion.

## 5. Conclusions

Across LBNL’s multi-site monitoring deployments and real-time data-processing systems, we have gained critical insights into induced seismicity at geothermal fields and developed scalable methods for seismic data acquisition, catalog generation, and analysis. This experience underscores the importance of long-term, high-resolution monitoring networks that are tailored to local geology and paired with automated detection pipelines. These systems lower magnitude completeness, improve seismic-risk mitigation, and support broader research objectives through public data access.

Accurate and timely seismic data collection is therefore foundational for both managing induced-seismicity risk and improving geothermal reservoir performance. As geothermal development scales toward deeper, hotter, and more complex formations, robust MEQ monitoring will be central to safe operations, reservoir management, and public acceptance of EGS projects.

**Acknowledgement**

We are grateful to the operators of each site for their collaboration in the deployment of seismometers. This work is supported by the U.S. Department of Energy's Hydrocarbons and Geothermal Energy Office under Contract No. DE-AC02-05CH11231 with Lawrence Berkeley National Laboratory. All seismic data in this paper are available at NSF SAGE (Table 1), and the earthquake catalogs, except the Geysers, are in the supplementary information in this manuscript.